\documentclass{article}[15pt]

\usepackage[dvips]{epsfig}
\usepackage{rotating}

\renewcommand{\baselinestretch}{1.25} 

\begin{document}
 
\large

\title{
A Complete Enumeration and Classification \\
of Two-Locus Disease Models\\
\vspace{0.2in}
\author{Wentian Li$^a$, Jens Reich $^b$ \\
{\small \sl a.Laboratory of Statistical Genetics, 
Rockefeller University, USA}\\
{\small \sl b. Department of Biomathematics, 
Max-Delbrück-Centrum, Berlin-Buch, Germany}\\
}\ 
\date{{\sl Human Heredity}, in press (1999)}
} 
\maketitle    
\markboth{\sl Li,neural net} {\sl Li, neural net}

{\bf Corresponding author:}  \\
Wentian Li, \\
Laboratory of Statistical Genetics, Box 192 \\
Rockefeller University,  1230 York Avenue \\
New York, NY 10021, USA. \\
Telephone: 212-327-7977 \\
FAX: 212-327-7996 \\
Email: wli@linkage.rockefeller.edu

\vspace{0.5in}

{\bf Key words: 

two-locus model, epistasis, 
identity-by-descent (IBD), correlation} 

\newpage

\begin{center}
{\bf ABSTRACT}
\end{center}

There are 512 two-locus, two-allele, two-phenotype, fully-penetrant disease 
models. Using the permutation between two alleles, between two loci, 
and between being affected and unaffected, one model can be considered 
to be equivalent to another model under the corresponding permutation. 
These permutations greatly reduce the number of two-locus models in 
the analysis of complex diseases. This paper determines the number of 
non-redundant two-locus models (which can be 102, 100, 96, 51, 50, 
or 48, depending on which permutations are used, and depending on 
whether zero-locus and single-locus models are excluded). Whenever
possible, these non-redundant two-locus models are classified
by their property. Besides the familiar features of multiplicative 
models (logical AND),  heterogeneity models (logical OR), and
threshold models, new classifications are added or expanded: modifying-effect 
models, logical XOR models, interference and negative interference
models (neither dominant nor recessive), conditionally dominant/recessive 
models, missing lethal genotype models, and highly symmetric models.  
The following aspects of two-locus models are studied: the marginal 
penetrance tables at both loci, the expected joint identity-by-descent
probabilities, and the correlation between marginal identity-by-descent
probabilities at the two loci. These studies are useful for linkage
analyses using single-locus models while the underlying disease model
is two-locus, and for correlation analyses using the linkage signals
at different locations obtained by a single-locus model.

\newpage

\section{Introduction}

Disease models involving two genes, usually called ``two-locus models"
(e.g. \cite{hogben,merry}), have been widely used in the study of 
complex diseases, including likelihood-based linkage analysis
\cite{green82,majumder89_1,lathrop90,schork93},
allele-sharing-based linkage analysis 
\cite{hodge81_1,dizier86,risch90_2,knapp94,cordell95,farrall97}, 
marker-association-segregation method \cite{clerget89,dizier94},
weighted-pairwise correlation method \cite{zinn98},
variance component analysis \cite{tiwari97_1,tiwari97_2,tiwari98},
recurrence risk of relatives \cite{trojak81,risch90_1,neuman92}, 
and segregation analysis 
\cite{green81,green84,green88,dizier89,dizier90,dizier93}.
Besides human genetics, two-locus models have also been used in the 
study of evolution, as well as genetic studies of inbreeding animals 
and plants.

Using two-locus models is a natural choice if the underlying disease 
mechanism indeed involves two or more genes, though there have been 
extensive discussions on the power of using single-locus models for 
linkage analysis in that situation \cite{green89,green90,goldin92,vieland92,goldin93,vieland93,sham94,schork94,olson95,dizier96,hodge97,todorov97}. 
Also, two-locus models have frequently been used in generating 
simulated datasets for testing various linkage methods and 
strategies \cite{goldgar92,davis96,falk97,todorov97,davis97,stoesz97,goldgar97,lucek98,badner98,green98,durner99}.
Although segregation analysis based on two-locus models is common
\cite{olson89,xu95,eccles97,scapoli97,juo98}, linkage analysis based
on two-locus models is relatively rare, due to the large number of combinations 
of two markers out of as many as 300 markers in the whole genome, 
due to the cost of a time-consuming calculation of the pedigree likelihood,
and due to a large number of possible possible interactions between
two genes.

One would naturally ask:  how many possible types of two-locus models exist? 
Complete enumerations and classifications of systems have been used in 
many other fields as a starting point of a study; for example, 
two-person two-move games in the study of game theory \cite{rapoport}, 
two-state three-input cellular automata in the study of dynamical systems 
\cite{wli90_2}, and two-symbol 3-by-3 lattice models in the study of protein 
folding \cite{hli96}. These types of studies lay out the space of all 
possibilities, with nothing missing. This paper follows a similar path 
in completely enumerating all two-locus two-allele two-phenotype disease models.

Strickberger \cite{strickberger} listed a few a types of two-locus models
encountered in experimental systems, though the number of phenotypes is
multiple (such as being a smooth, partly rough and fully rough Mendelian 
pea), instead of binary (such as affected and unaffected). 
Defrise-Gussenhoven \cite{defrise} listed five types of two-locus
models, which were followed up by a study by Greenberger \cite{green81}. 
Neuman and Rice listed six two-locus models \cite{neuman92}.
Nevertheless, nobody provided a complete list of all possible
two-locus models.

This complete enumeration of all two-locus models can be useful
when a linkage signal is observed in two separated regions, or if 
two candidate genes with known locations are studied.  In 
these situations, it is of interest to determine the nature of 
the interaction between the two disease genes (e.g. \cite{dizier94}). 
Without knowing all possible forms of interaction, such determination 
is not complete.

A list of all two-locus models is perhaps useful for
likelihood-based linkage analysis, but may not be essential.
In such a linkage analysis, parameters in the two-locus model
can be determined by a maximum likelihood method, and the fitted
values are generally continuous rather than discrete. The enumeration
of two-locus models in this paper, however, uses discrete parameter
values. Nevertheless, during the stage of interpretation of
the result, the classification of two-locus models discussed in
section 3 can be useful.

Since most likelihood-based linkage analyses still use
single-locus disease models, it is of interest to know how
closely a single-locus model approximates a two-locus model.
For this purpose, we examine the marginal penetrance (on both loci)
of all two-locus models, which should be the optimal parameter
value if a single-locus model is used for the linkage analysis
\cite{sham94}. The question of which two-locus models can be 
reasonably approximated by single-locus models, or which two-locus
interaction can be detected by single-locus linkage analysis,
can be easily answered by this marginal penetrance information.
This topic will be discussed in section 4.

Allele-sharing-based linkage analysis requires a calculation
of the expected allele sharing between a relative pair 
under a certain disease model \cite{dizier86,risch90_2,knapp94,cordell95,farrall97}.
We provide a new formulation for this calculation which is an
extension of the classical Li-Sacks method \cite{lisacks,ccli_book},
which in turn is based on the Bayes' theorem. This topic will be
discussed in section 5.

It has been suggested that interaction or epistasis between two
regions can be detected by calculating the correlation between
two linkage signals, each determined by a single-locus
linkage analysis \cite{maclean,cox}. A positive correlation 
may suggest interaction (epistasis), and a negative correlation
may suggest heterogeneity \cite{maclean,cox}. We examine such
a correlation for all two-locus models, which not only confirms 
this simple rule-of-thumb, but also generalizes to other two-locus
models. This topic will be discussed in section 6.

\section{Enumeration of two-locus models}

A two-locus model is typically represented by a 3-by-3 penetrance table.
The row label gives the three possible genotypes of the first disease
locus (i.e. aa,aA,AA, where A might be considered as the disease allele at
locus 1), and the column label gives the genotypes for the second locus
(i.e. bb,bB,BB, where B is the disease allele at locus 2):
\begin{equation}
\label{eq1}
\{ f_{ij} \} =
\begin{array}{cccc}
 &bb &bB & BB\\
aa & f_{11} & f_{12} & f_{13} \\
aA & f_{21} & f_{22} & f_{23} \\
AA & f_{31} & f_{32} & f_{33} \\
\end{array}
\end{equation}
The table element $f_{ij}$ (``penetrance") is the probability 
of being affected with the disease when the genotype at the first 
locus is $i$, and that of the second locus is $j$.  In the most general 
case,  $f_{ij}$'s range from 0 to 1. Models defined on continuously
varying parameters are hard to be classified to a few discrete categories. 
On the other hand, if the the allowed values of $f_{ij}$'s are 0 and 1 
only (``fully penetrant"), we can categorize the 
nine-parameter space to $2^9=512$ distinct points. We use the
following notation to label each of these 512 fully-penetrant two-locus
models:
\begin{equation}
\mbox{``model number"$_{10}$} =
(f_{11} f_{12} f_{13} f_{21} f_{22} f_{23} f_{31} f_{32} f_{33})_2
\end{equation}
where the subscript of 2 or 10 indicates whether the number is represented
as binary or decimal. For example, if a model has $f_{13}=1$ 
and other $f_{ij}$'s are zero,  the binary representation of the 
penetrance table is (001000000)$_2$, which is 64 in decimal notation,
or model M64. Model numbers range from 0 to 511.

The number of non-redundant two-locus models is less than 512 due to the 
following considerations: (i) if all $f_{ij}$'s are 0 (or 1), the model is 
a zero-locus model; (ii) if the elements of the penetrance table do not 
change with row (or with column), it is a single-locus model;
the nature of the model should not change (iii) if the first and second 
locus are exchanged;  (iv) if the two alleles in the first
(or second) locus are exchanged; or (v) if the affection status is exchanged.
We will show below that when the symmetries implied by permutation 
(iii) and (iv) are imposed, the number of non-redundant two-locus model 
($N_1$) is 102; when (iii),(iv),(v) are considered, 
the number ($N_2$) is 51. Subtracting zero-locus and/or single-locus
models, we get $N_1-$2=100, $N_1-$6=96, $N_2-$1=50, and $N_2-$3=48.

This result of the number of non-redundant two-locus models is based on 
the counting theorem by P\'{o}lya and de Bruijn \cite{polya,debruijn}.
Cotterman pioneered combinatorial genetics, but he only enumerated
single-locus multiple-allele models \cite{cotterman}.  Although 
Hartle and Maruyama had already applied the counting theorem 
to enumerate genetic models \cite{hartl}, we would like to repeat
and simplify the derivation to focus on our particular case, i.e., the 
two-locus two-allele models. 

To do so, it is necessary to review the concept of ``cycle index" below. 
 If a permutation is applied to a set of $m$ elements, some 
elements are invariant under this permutation 
($b_1$ of them), some form cycles of length 2 ($b_2$ of them), 
some form cycles of length 3 ($b_3$ of them), etc. For each 
permutation, construct a polynomial with $m$ variables:
$$
        x_1^{b_1} x_2^{b_2} x_3^{b_3} \cdots x_m^{b_m}.
$$
Going through all permutation $p$'s that are part of the permutation 
group $P$ (suppose the number of permutations is $|P|$), the cycle 
index is defined as the polynomial:
$$
C(x_1, x_2, \cdots x_m) \equiv
\frac{1}{|P|}
\sum_{ p \in P} x_1^{b_1} x_2^{b_2} x_3^{b_3} \cdots x_m^{b_m}.
$$

For two-locus models, there are 9 genotypes, and  eight permutations
can be considered on this set of genotypes: (i) the identity operation; 
(ii) exchange alleles $a$ and $A$; (iii) exchange alleles $b$ and 
$B$; (iv) exchange the  first and the second locus; (v) is (ii) 
plus (iii); (vi) is (ii) plus (iv); (vii) is (iii) plus (iv); 
(viii) is (v) plus (iv). The cycle index for this group of eight 
permutations on the 9 genotypes is:
$$
C_{geno}(x_1, x_2, \cdots x_9) =
\frac{x_1^9 +4 x_1^3 x_2^3 +x_1 x_2^4 + 2 x_1 x_4^2}{8}.
$$

By P\'{o}lya's counting theorem (theorem 5.1 in \cite{debruijn})
the number of non-redundant two-locus models, without considering
permutations in phenotype, is equal to the cycle index of the
permutation group on the genotype evaluated by replacing all variables 
by the number of phenotypes (which is 2), i.e.: 
$$
N_1 = \frac{2^9 + 2^8 + 2^5 +2^4}{8} =102.
$$

When all 0's in the penetrance table are switched to 1 and 1's switched
to 0, one two-locus model becomes another two-locus model. If we consider 
these two models as equivalent, the number of non-redundant models is
$$
N_2 = \frac{N_1}{2} = 51.
$$
Actually, the same conclusion can be obtained by considering not only
the cycle index of the permutation group on the genotype, but also
that of a permutation group on the phenotype, then using de Bruijn's 
generalization of P\'{o}lya's theorem (see Appendix 1).  The advantage 
of this approach is that if a more complicated permutation group 
applied to phenotype is considered, the method to get $N_2$ by a simple
division of $N_1$ would not work.

\section{Classifying two-locus models}

This section discusses some possible classification schemes of
two-locus models. No attempt is made to exhaustively classify
all models, considering the fact that some ``exotic" models can
never be classified using familiar terms. What we have here
is a collection of classification schemes, each selecting a subset
of models by a special property they possess. As a comparison,
out of the 50 models listed in this paper,  Defrise-Gussenhoven
studied M1, M3, M11, M15, M27 \cite{defrise}; Greenberg studied
M1, M3, M27 \cite{green81}; and Neuman and Rice studied M1, M3, M11, 
M15, M27,M78 \cite{neuman92}. All $N_2-1=$50 models 
are listed in Table 1. The $N_1-N_2-1=$50 models generated by 
switching affecteds and unaffecteds (plus possibly other permutations
between loci and allele) are listed in Table 2 for convenience.

\begin{table}
\begin{tabular}{|ccc|c|ccc|c|ccc|c|ccc|c|ccc|c|ccc|c|ccc|}
\multicolumn{4}{c}{{\bf M1(RR)}} & \multicolumn{4}{c}{M2} & 
	\multicolumn{4}{c}{{\bf M3(RD)}} & \multicolumn{4}{c}{M5} & 
	\multicolumn{4}{c}{{\bf M7(1L:R)}} & \multicolumn{4}{c}{M10} &
	\multicolumn{3}{c}{{\bf M11 (T)}}\\
\cline{1-3} \cline{5-7} \cline{9-11} \cline{13-15} \cline{17-19}
\cline{21-23} \cline{25-27}
0&0&0 & &0&0&0& &0&0&0& &0&0&0& &0&0&0&  &0&0&0&  &0&0&0\\
0&0&0 & &0&0&0& &0&0&0& &0&0&0& &0&0&0&  &0&0&1&  &0&0&1 \\
0&0&1 & &0&1&0& &0&1&1& &1&0&1& &1&1&1&  &0&1&0&  &0&1&1 \\
\cline{1-3} \cline{5-7} \cline{9-11} \cline{13-15} \cline{17-19}
\cline{21-23} \cline{25-27}

\multicolumn{4}{c}{M12} & \multicolumn{4}{c}{M13} & 
	\multicolumn{4}{c}{M14} & \multicolumn{4}{c}{{\bf M15(Mod)}} & 
	\multicolumn{4}{c}{M16} & \multicolumn{4}{c}{M17} &
	\multicolumn{3}{c}{M18}\\
\cline{1-3} \cline{5-7} \cline{9-11} \cline{13-15} \cline{17-19}
\cline{21-23} \cline{25-27}
0&0&0& &0&0&0& &0&0&0& &0&0&0 & &0&0&0& &0&0&0& &0&0&0 \\
0&0&1& &0&0&1& &0&0&1& &0&0&1 & &0&1&0& &0&1&0& &0&1&0 \\
1&0&0& &1&0&1& &1&1&0& &1&1&1 & &0&0&0& &0&0&1& &0&1&0\\
\cline{1-3} \cline{5-7} \cline{9-11} \cline{13-15} \cline{17-19}
\cline{21-23} \cline{25-27}

\multicolumn{4}{c}{M19} & \multicolumn{4}{c}{M21} & 
\multicolumn{4}{c}{M23} & \multicolumn{4}{c}{M26} & 
\multicolumn{4}{c}{{\bf M27 (DD)}} & \multicolumn{4}{c}{M28} &
\multicolumn{3}{c}{M29}\\
\cline{1-3} \cline{5-7} \cline{9-11} \cline{13-15} \cline{17-19}
\cline{21-23} \cline{25-27}
0&0&0&  &0&0&0 & &0&0&0& &0&0&0& &0&0&0& &0&0&0& &0&0&0 \\
0&1&0&  &0&1&0 & &0&1&0& &0&1&1& &0&1&1& &0&1&1& &0&1&1 \\
0&1&1&  &1&0&1 & &1&1&1& &0&1&0& &0&1&1& &1&0&0& &1&0&1\\
\cline{1-3} \cline{5-7} \cline{9-11} \cline{13-15} \cline{17-19}
\cline{21-23} \cline{25-27}

\multicolumn{4}{c}{M30} & \multicolumn{4}{c}{M40} & \multicolumn{4}{c}{M41} &
\multicolumn{4}{c}{M42} & \multicolumn{4}{c}{M43} & \multicolumn{4}{c}{M45} &
\multicolumn{3}{c}{{\bf M56(1L:I)}}\\
\cline{1-3} \cline{5-7} \cline{9-11} \cline{13-15} \cline{17-19}
\cline{21-23} \cline{25-27}
 0&0&0& &0&0&0& &0&0&0& &0&0&0& &0&0&0 & &0&0&0& &0&0&0 \\
 0&1&1& &1&0&1& &1&0&1& &1&0&1& &1&0&1 & &1&0&1& &1&1&1 \\
 1&1&0& &0&0&0& &0&0&1& &0&1&0& &0&1&1 & &1&0&1& &0&0&0 \\
\cline{1-3} \cline{5-7} \cline{9-11} \cline{13-15} \cline{17-19}
\cline{21-23} \cline{25-27}

\multicolumn{4}{c}{M57} & \multicolumn{4}{c}{M58} & \multicolumn{4}{c}{M59} &
\multicolumn{4}{c}{M61} & \multicolumn{4}{c}{M68} & \multicolumn{4}{c}{M69} &
\multicolumn{3}{c}{M70}\\
\cline{1-3} \cline{5-7} \cline{9-11} \cline{13-15} \cline{17-19}
\cline{21-23} \cline{25-27}
0&0&0& &0&0&0& &0&0&0 & &0&0&0& &0&0&1& &0&0&1& &0&0&1 \\
1&1&1& &1&1&1& &1&1&1 & &1&1&1& &0&0&0& &0&0&0& &0&0&0 \\
0&0&1& &0&1&0& &0&1&1 & &1&0&1& &1&0&0& &1&0&1& &1&1&0 \\
\cline{1-3} \cline{5-7} \cline{9-11} \cline{13-15} \cline{17-19}
\cline{21-23} \cline{25-27}

\multicolumn{4}{c}{{\bf M78(XOR)}} & \multicolumn{4}{c}{M84} & 
\multicolumn{4}{c}{M85} & \multicolumn{4}{c}{M86} & 
\multicolumn{4}{c}{M94} & \multicolumn{4}{c}{M97} &
\multicolumn{3}{c}{M98}\\
\cline{1-3} \cline{5-7} \cline{9-11} \cline{13-15} \cline{17-19}
\cline{21-23} \cline{25-27}
0&0&1 & &0&0&1& &0&0&1& &0&0&1& &0&0&1& &0&0&1 & &0&0&1 \\
0&0&1 & &0&1&0& &0&1&0& &0&1&0& &0&1&1& &1&0&0 & &1&0&0 \\
1&1&0 & &1&0&0& &1&0&1& &1&1&0& &1&1&0& &0&0&1 & &0&1&0 \\
\cline{1-3} \cline{5-7} \cline{9-11} \cline{13-15} \cline{17-19}
\cline{21-23} \cline{25-27}

\multicolumn{4}{c}{M99} & \multicolumn{4}{c}{M101} & 
\multicolumn{4}{c}{M106}& \multicolumn{4}{c}{M108} & 
\multicolumn{4}{c}{M113} & \multicolumn{4}{c}{M114} &
\multicolumn{3}{c}{M170} \\
\cline{1-3} \cline{5-7} \cline{9-11} \cline{13-15} \cline{17-19}
\cline{21-23} \cline{25-27}
0&0&1& &0&0&1& &0&0&1& &0&0&1 & &0&0&1& &0&0&1& &0&1&0 \\
1&0&0& &1&0&0& &1&0&1& &1&0&1 & &1&1&0& &1&1&0& &1&0&1 \\
0&1&1& &1&0&1& &0&1&0& &1&0&0 & &0&0&1& &0&1&0& &0&1&0 \\
\cline{1-3} \cline{5-7} \cline{9-11} \cline{13-15} \cline{17-19}
\cline{21-23} \cline{25-27}

\multicolumn{3}{c}{M186} \\
\cline{1-3}
0&1&0 \\
1&1&1 \\
0&1&0 \\
\cline{1-3}

\end{tabular}
\caption{
The penetrance tables of all $N_2-1=$50 two-locus models. Each model 
represents a group of equivalent models under permutations. The 
representative model is the one with the smallest model number. The six 
models studied in Neuman and Rice (``RR,RD,DD,T,Mod,XOR") \cite{neuman92}, 
as well as two single-locus models (``1L") -- the recessive (R) and
the interference (I) model, are marked.
}
\end{table}

\begin{table}
\footnotesize
\begin{tabular}{|ccc|c|ccc|c|ccc|c|ccc|c|ccc|c|ccc|c|ccc|c}
\multicolumn{4}{c}{M31$\rightarrow$15 {\bf (Mod)}} & 
\multicolumn{4}{c}{M47$\rightarrow$23} & 
\multicolumn{4}{c}{M63$\rightarrow$7{\bf (1L:D)}} &
\multicolumn{4}{c}{M71$\rightarrow$59} & 
\multicolumn{4}{c}{M79$\rightarrow$27{\bf (R+R)}} & 
\multicolumn{4}{c}{M87$\rightarrow$46} &
\multicolumn{4}{c}{M95$\rightarrow$11{\bf (T)}}\\
\cline{1-3} \cline{5-7} \cline{9-11} \cline{13-15} \cline{17-19}
\cline{21-23} \cline{25-27}
0&0&0 & &0&0&0& &0&0&0& &0&0&1& &0&0&1&  &0&0&1&  &0&0&1\\
0&1&1 & &1&0&1& &1&1&1& &0&0&0& &0&0&1&  &0&1&0&  &0&1&1 \\
1&1&1 & &1&1&1& &1&1&1& &1&1&1& &1&1&1&  &1&1&1&  &1&1&1 \\
\cline{1-3} \cline{5-7} \cline{9-11} \cline{13-15} \cline{17-19}
\cline{21-23} \cline{25-27}

\multicolumn{4}{c}{M102$\rightarrow$94} & 
\multicolumn{4}{c}{M103$\rightarrow$30} & 
\multicolumn{4}{c}{M105$\rightarrow$61} &
\multicolumn{4}{c}{M107$\rightarrow$29} &
\multicolumn{4}{c}{M109$\rightarrow$57} &
\multicolumn{4}{c}{M110$\rightarrow$86} &
\multicolumn{4}{c}{M111$\rightarrow$19}\\
\cline{1-3} \cline{5-7} \cline{9-11} \cline{13-15} \cline{17-19}
\cline{21-23} \cline{25-27}
0&0&1& &0&0&1& &0&0&1& &0&0&1 & &0&0&1& &0&0&1& &0&0&1 \\
1&0&0& &1&0&0& &1&0&1& &1&0&1 & &1&0&1& &1&0&1& &1&0&1 \\
1&1&0& &1&1&1& &0&0&1& &0&1&1 & &1&0&1& &1&1&0& &1&1&1\\
\cline{1-3} \cline{5-7} \cline{9-11} \cline{13-15} \cline{17-19}
\cline{21-23} \cline{25-27}

\multicolumn{4}{c}{M115$\rightarrow$99} & 
\multicolumn{4}{c}{M117$\rightarrow$106} &
\multicolumn{4}{c}{M118$\rightarrow$78} &
\multicolumn{4}{c}{M119$\rightarrow$14} & 
\multicolumn{4}{c}{M121$\rightarrow$45} & 
\multicolumn{4}{c}{M122$\rightarrow$101} &
\multicolumn{4}{c}{M123$\rightarrow$13}\\
\cline{1-3} \cline{5-7} \cline{9-11} \cline{13-15} \cline{17-19}
\cline{21-23} \cline{25-27}
0&0&1&  &0&0&1 & &0&0&1& &0&0&1& &0&0&1& &0&0&1& &0&0&1 \\
1&1&0&  &1&1&0 & &1&1&0& &1&1&0& &1&1&1& &1&1&1& &1&1&1 \\
0&1&1&  &1&0&1 & &1&1&0& &1&1&1& &0&0&1& &0&1&0& &0&1&1\\
\cline{1-3} \cline{5-7} \cline{9-11} \cline{13-15} \cline{17-19}
\cline{21-23} \cline{25-27}

\multicolumn{4}{c}{M124$\rightarrow$108} & 
\multicolumn{4}{c}{M125$\rightarrow$41} & 
\multicolumn{4}{c}{M126$\rightarrow$70} &
\multicolumn{4}{c}{M127$\rightarrow$3{\bf (D+R)}} & 
\multicolumn{4}{c}{M171$\rightarrow$85} & 
\multicolumn{4}{c}{M173$\rightarrow$113} &
\multicolumn{4}{c}{M175$\rightarrow$21}\\
\cline{1-3} \cline{5-7} \cline{9-11} \cline{13-15} \cline{17-19}
\cline{21-23} \cline{25-27}
 0&0&1& &0&0&1& &0&0&1& &0&0&1& &0&1&0 & &0&1&0& &0&1&0 \\
 1&1&1& &1&1&1& &1&1&1& &1&1&1& &1&0&1 & &1&0&1& &1&0&1 \\
 1&0&0& &1&0&1& &1&1&0& &1&1&1& &0&1&1 & &1&0&1& &1&1&1 \\
\cline{1-3} \cline{5-7} \cline{9-11} \cline{13-15} \cline{17-19}
\cline{21-23} \cline{25-27}

\multicolumn{4}{c}{M187$\rightarrow$69} & 
\multicolumn{4}{c}{M189$\rightarrow$97} & 
\multicolumn{4}{c}{M191$\rightarrow$5} &
\multicolumn{4}{c}{M229$\rightarrow$114} & 
\multicolumn{4}{c}{M231$\rightarrow$28} & 
\multicolumn{4}{c}{M238$\rightarrow$84} &
\multicolumn{4}{c}{M239$\rightarrow$17}\\
\cline{1-3} \cline{5-7} \cline{9-11} \cline{13-15} \cline{17-19}
\cline{21-23} \cline{25-27}
0&1&0& &0&1&0& &0&1&0 & &0&1&1& &0&1&1& &0&1&1& &0&1&1 \\
1&1&1& &1&1&1& &1&1&1 & &1&0&0& &1&0&0& &1&0&1& &1&0&1 \\
0&1&1& &1&0&1& &1&1&1 & &1&0&1& &1&1&1& &1&1&0& &1&1&1 \\
\cline{1-3} \cline{5-7} \cline{9-11} \cline{13-15} \cline{17-19}
\cline{21-23} \cline{25-27}

\multicolumn{4}{c}{M245$\rightarrow$98} & 
\multicolumn{4}{c}{M247$\rightarrow$12} & 
\multicolumn{4}{c}{M254$\rightarrow$68} &
\multicolumn{4}{c}{M255$\rightarrow$1{\bf (D+D)}} & 
\multicolumn{4}{c}{M325$\rightarrow$186} & 
\multicolumn{4}{c}{M327$\rightarrow$58} &
\multicolumn{4}{c}{M335$\rightarrow$26}\\
\cline{1-3} \cline{5-7} \cline{9-11} \cline{13-15} \cline{17-19}
\cline{21-23} \cline{25-27}
0&1&1 & &0&1&1& &0&1&1& &0&1&1& &1&0&1& &1&0&1 & &1&0&1 \\
1&1&0 & &1&1&0& &1&1&1& &1&1&1& &0&0&0& &0&0&0 & &0&0&1 \\
1&0&1 & &1&1&1& &1&1&0& &1&1&1& &1&0&1& &1&1&1 & &1&1&1 \\
\cline{1-3} \cline{5-7} \cline{9-11} \cline{13-15} \cline{17-19}
\cline{21-23} \cline{25-27}

\multicolumn{4}{c}{M341$\rightarrow$170} & 
\multicolumn{4}{c}{M343$\rightarrow$42} & 
\multicolumn{4}{c}{M351$\rightarrow$10} &
\multicolumn{4}{c}{M365$\rightarrow$56{\bf (1L:$\overline{I}$)}} & 
\multicolumn{4}{c}{M367$\rightarrow$18} & 
\multicolumn{4}{c}{M381$\rightarrow$40} &
\multicolumn{4}{c}{M383$\rightarrow$2} \\
\cline{1-3} \cline{5-7} \cline{9-11} \cline{13-15} \cline{17-19}
\cline{21-23} \cline{25-27}
1&0&1& &1&0&1& &1&0&1& &1&0&1 & &1&0&1& &1&0&1& &1&0&1 \\
0&1&0& &0&1&0& &0&1&1& &1&0&1 & &1&0&1& &1&1&1& &1&1&1 \\
1&0&1& &1&1&1& &1&1&1& &1&0&1 & &1&1&1& &1&0&1& &1&1&1 \\
\cline{1-3} \cline{5-7} \cline{9-11} \cline{13-15} \cline{17-19}
\cline{21-23} \cline{25-27}

\multicolumn{3}{c}{M495$\rightarrow$16} \\
\cline{1-3}
1&1&1 \\
1&0&1 \\
1&1&1 \\
\cline{1-3}

\end{tabular}
\caption{
The penetrance tables of $N_1-N_2-1=$50 two-locus models. These models
are equivalent to the models in Table 1 by the  $0 \leftrightarrow 1 $ 
permutation plus possibly other permutations between two loci
and between two alleles. The most familiar models, including the
two single-locus models -- the dominant (D) and the
negative interference ($\overline{I}$) model, are marked.
}
\end{table}

\large

We first review the 6 models studied in \cite{neuman92}:
\begin{enumerate}
\item
{\bf Jointly-recessive-recessive model (RR)}

M1 requires two copies of the disease alleles from {\sl both}
loci to be affected. This model was studied as early as 1952
\cite{steinberg,ccli,majumder89_2}, and can also be called 
``recessive complementary".

\item
{\bf Jointly-dominant-dominant model (DD)}

M27 requires at least one copy of the disease allele from {\sl both}
loci to be affected. This model can also be called ``dominant 
complementary".

\item
{\bf Jointly-recessive-dominant model (RD)}

M3 requires two copies of disease alleles from the first locus and 
at least one disease allele from the second locus to be affected.

Note that the {\bf Heterogeneity models} (logical OR models)
discussed in \cite{neuman92} are equivalent to the above three 
RR, DD, RD models by the 0 $\leftrightarrow$ 1 permutation in 
the penetrance table plus possibly some permutations between two loci
and/or two alleles.  RR model becomes D+D model, DD model 
becomes R+R, and RD becomes D+R \cite{durner92}.

\item
{\bf A modifying-effect model (Mod)}

M15  can be modified to a single-locus recessive model if the
penetrance at the genotype aA-BB is changed from 1 to 0.
This model is one of the ``modifying-effect models" and
``almost single-locus models" discussed below.

\item
{\bf Threshold model (T)}

M11 requires at least three disease alleles, regardless of which 
locus the disease alleles are from, to be affected. M95, which
is equivalent to M11, requires at least two disease alleles
to be affected. 

\item
{\bf An exclusive OR model (XOR)}

M78 is almost the R+R model except for the two-locus genotype
AA-BB. This model was used to model the genetics of handedness 
\cite{levy}. In fact, M78 is one of the ``exclusive OR" models
to be discussed below.
\end{enumerate}

There are also the following classification schemes 
\begin{itemize}
\item
{\bf Single-locus models (1L)}:

M7 is a single-locus recessive model (it is also equivalent to
a single-locus dominant model M63, by 0 $\leftrightarrow$ 1 
permutation in the penetrance table, followed by a permutation 
between alleles $a$ and $A$).
M56 is a single-locus ``interference" (the term used by Johnson is
``metabolic interference"  \cite{johnson}), or ``maximum heterozygosity 
model". As discussed in details by Johnson \cite{johnson}, in this 
hypothetical model, neither allele $a$ nor $A$ is really abnormal;
only when the gene products interact, can there be harmful effects. 
M365 is equivalent to M56 by the  0 $\leftrightarrow$ 1 permutation
(plus a permutation between two loci),
which can be called a ``negative interference model" or a ``maximum 
homozygosity model". Models similar to M56 and M365, which are
neither dominant nor recessive,  will be discussed more below. 
M7,M63,M56,M365 are labeled as $R$,$D$,$I$, $\overline{I}$.

We can classify two-locus models which are one-mutation away
from single-locus models as {\bf almost single-locus models}. The
modifying-effect model M15 is actually an almost single-locus
model. Others include M23, M57, M58 ( 0 $\rightarrow$ 1
mutation in the penetrance table), M3, M5, M59, and M61
(1 $\rightarrow$ 0 mutation in the penetrance table).

\item
{\bf Logical AND (multiplicative) models}:

The logical AND operation on two binary variables is defined
as: 0 AND 0 = 0, 0 AND 1=0, 1 AND 0 = 0, 1 AND 1 =1. Imagine
that the penetrance table receives a contribution from both loci, 
$\{ g1_i \}$ and $\{ g2_j \}$ ($i,j=1,2,3$), and the penetrance 
value can be represented as a product of the two contributions
\cite{moran}:
$$
f_{ij} = g1_i \mbox{ AND } g2_j,
$$
This class of model includes M1(RR), M2(RI), M3(RD), M5(R$\overline{I}$), 
M16(II), M18(DI), M27(DD), M40(I$\overline{I}$), M45(D$\overline{I}$), 
and M325 ($\overline{I}\overline{I}$), where $R,D,I,\overline{I}$ are 
dominant, recessive, interference, and negative interference single-locus
models.  M325 is equivalent to M186 by the permutation in the affection 
status. Although M7 and M56 are also logical AND models, they are actually 
trivial single-locus models. One can see that for M45, for example,
when the second and third columns in the penetrance table
are switched, all non-zero elements form a rectangular block. It is
true for any multiplicative model that such a rectangular block can
be formed by switching columns and/or rows.

The special interest of multiplicative models lies in the
fact that the probability of the value of identity-by-descent 
at one locus is independent of the other locus \cite{hodge81_1}.
In other words, if one uses the joint identity-by-descent between
affected sibpairs to study a possible interaction between
two locations, such an interaction cannot be detected. More
on the calculation of the  probability of identity-by-descent 
values will be discussed below.

\item
{\bf Logical OR (heterogeneity) models}:

The logical OR operation on two binary variables is defined
as: 0 OR 0 = 0, 0 OR 1 = 1, 1 OR 0 = 1, 1 OR 1 = 1. The
0 $\leftrightarrow$ 1 permutation in the penetrance table will
transform a logical AND model to a {\bf logical OR model},
or a {\bf heterogeneity model}.  Note that for fully-penetrant models, 
we cannot have an exact, but only approximate, {\bf additive models} 
in the original sense, since 1+1=2 is larger than what is allowed 
by a penetrance.

\item
{\bf Logical XOR models}:

The logical XOR (exclusive OR) operation on two binary variables
is defined as: 0 XOR 0 =0, 0 XOR 1=1, 1 XOR 0=1, 1 XOR 1=0.
The last equation makes XOR an extremely non-linear operation.
Because of this property, XOR is a favorite function to illustrate 
the advantage of artificial neural networks over linear discrimination 
and linear regression (e.g. \cite{bishop}). Logical XOR
two-locus models include M78 (as discussed earlier), M113, and
M170.

\item
{\bf Conditional dominant (recessive) models}:

These are models where the first (or the second) locus behaves like
a dominant (or recessive) model if the second (or the first) locus
takes a certain genotype.  For example, the first locus in M11 
behaves as a recessive model when the genotype at the second 
locus is $bB$, but as a dominant model when the genotype 
at the second locus is $BB$. Models similar to M11 include: 
M1(RR), M2, M3(DR), M5, M13, M15(Mod), M18, M19, M23, and M45.

\item
{\bf Interference models: neither dominant nor recessive}:

We can extend the single-locus ``neither dominant nor recessive"
models M56 and M365 to two-locus models. In positive interferences, 
two otherwise normal proteins produced at two loci interact to lead 
to the disease. In negative interferences, two 
complementary proteins lead to a functional product and an unaffected 
person, whereas the lack of either complementary component leads to
affection. These following models illustrate the situation: 
M68, M186, and M170.

In M68, the only two-locus genotypes that lead to the disease
are aa-BB and bb-AA. Suppose an abnormal effect is caused by
an interaction between the protein product generated from allele $a$
and  that from $B$, or between the protein products from $b$ and $A$.
Then only the above two two-locus genotypes lead to the maximum
abnormal effect. This model was studied in \cite{merry2}.

For M325, which is equivalent to M186 by the 0 $\leftrightarrow$ 1 
permutation in the penetrance table, four two-locus genotypes lead
to the disease: aa-bb, aa-BB, AA-bb, AA-BB. This is a situation
where maximum doses of the protein produced at both loci lead to 
the disease. From this perspective, M325 is a ``maximum 
homozygosity" model (and M186 a ``maximum heterozygosity" model).

For M170, four two-locus genotypes lead to the disease: aa-bB, 
aA-bb, aA-BB, AA-bB. The difference between M170 and M186 is that
the double-heterozygosity genotype aA-bB does not lead to the disease,
whereas all other heterozygous genotypes lead to the disease.
One might consider that there is another between-locus interference 
besides the within-locus interference, and the two interferences
cancel out.

In {\sl Drosophila} genetics, the phenomenon of metabolic
interference is called ``negative complementation"
\cite{welshons, wilkie}. For example, the Notch gene has two
types, ``enhancers" and ``suppressors". The homozygotes for
both types are viable, whereas the heterozygotes are lethal.

The phenomenon of ``maternal-fetal incompatibility" \cite{ober}
is reminiscent of, but not identical to, the interference
we discuss here. This incompatibility is between the red blood cells 
in the mother and in the fetus, due to the inheritance of two 
different alleles from the mother and the father. This occurs only
if the fetus' genotype is heterozygous.

\item
{\bf More modifying-effect models}:

Just as M15 is a modified version of the single-locus recessive 
model, any model whose penetrance table is one mutation away from 
a classified model has a modifying-effect on the latter. For example, 
changing the penetrance value from 1 to 0 in M41 at the two-locus 
genotype aA-bb makes it a single-locus dominant model. Other 
modifying-effect models are listed in Table 3.

\item
{\bf Missing lethal genotype models}:

We consider the following situation: a genetic disease requires a 
minimum number of disease alleles from either/both locus/loci (i.e. alleles 
A and B), which lead to models similar to the threshold model (M11 or 
its equivalent model M95).  Nevertheless, if the disease is lethal, all 
individuals carrying a large number of disease alleles disappear from the
population. Consequently, it is impossible to have the two-locus genotype 
with the maximum number of disease alleles (e.g. AA-BB, AA-bB, aA-BB). 
Although all possible two-locus genotypes are specified in the penetrance 
table, some genotypes never appear in the population. Effectively, we 
may replace the penetrances at these genotypes by ``not available"
+'s or 0's.

For example, in the penetrance table below, the AA-BB genotype
is missing from the population, thus its penetrance is replaced
by a ``+":
\begin{equation}
\begin{array}{cccc}
 &bb &bB & BB\\
aa & 0 & 0 & 0 \\
aA & 0 & 0 & 1 \\
AA & 0 & 1 & + \\
\end{array}
\end{equation}
Since we will never have a chance to use the penetrance represented
by +, it might be replaced by a 0, and  become model M10. The 
following models also belong to this class: M2, M12, M14,
M18,  M26, M28, M30, M78, M84, M86, M94, M124 (equivalent to
M108),  M126 (equivalent to M70), M254 (equivalent to M68) 
(the +'s appear in the lower-right corner), M3, M19 (the +'s 
appear in the upper-right corner).  A model similar to M84 was 
discussed in \cite{frankel}. 

The discussion presented here illustrates a general principle:
even if two two-locus models may differ in their penetrance
table, they can be effectively identical if the differing element
appears with a very small probability.

\item
{\bf Highly symmetric models}:
 
During the discussion of P\'{o}lya's theorem, eight permutations
were listed including the identity operation and seven other 
permutations. Whether a model is invariant or not under the 
seven permutations provides a measure of the degree 
of symmetry of the model. For example, M40 is invariant 
under three permutations:  exchange of alleles $a$ and $A$, 
exchange of alleles $b$ and $B$, exchange of both $a$, $A$, and $b$, $B$.
Other models which are invariant under a large number of permutations
(indicated by the number in the parentheses) include: M16 (7), M40 (3),
 M68 (3), M84 (3), M170 (7), M186 (7). M56 is excluded 
because it is a single-locus model.

Models that are symmetric with respect to permutation of two loci
need only one single-locus model to approximate both loci. Models
that are symmetric with respect to permutation of two alleles
might be more relevant to common diseases.

\end{itemize}

Admittedly, there are ``exotic" models which have yet to be classified. 
Although one can relax the definitions of modifying-effect and 
interference models to incorporate them, they are less likely to 
be useful in modeling the gene-gene interaction in real situations. Table 3
summarizes what we have discussed in this section.

\begin{table}
\begin{center}
\begin{tabular}{| c | l | c | l |}

\hline
model & classifications &model & classifications \\
\hline
{\bf M1} & {\bf RR},C,AND,$S_{L}$,[3,68] (M255 $\rightarrow$ {\bf D+D},OR)
	&M43& [11] \\
M2 & L,C,AND,$S_A$, [3] 
	&M45 & C,AND,$S_A$ \\     
{\bf M3} & L, {\bf RD}, C, AND, [1,7,11] (M127 $\rightarrow$ {\bf D+R},OR)
	&{\bf M56} & {\bf 1L:I}, $S_{A,AA}$ 
	(M365 $\rightarrow$ {\bf 1L:$\overline{I}$}) \\
M5 & C, AND, $S_A$, [1,7]    
	&M57  & [56]            \\
{\bf M7} & {\bf 1L:R}, $S_A$, [3] (M63 $\rightarrow$ {\bf 1L:D})
        &M58 & $S_A$, [56,186]   \\
M10 & L, $S_L$, [11]  
	&M59 & [27] (M71 $\rightarrow$ [7])   \\
{\bf M11} & {\bf T}, C, $S_L$, [3,27]
        &M61 & $S_A$ (M105 $\rightarrow$ [7])  \\
M12 & L,[1] 
	&{\bf M68} & I, $S_{L,AA}$, [1] (M254 $\rightarrow$ L) \\
M13 & C, [3]             
	&M69 & $S_L$, [68] (M187 $\rightarrow$ [186]) \\
M14 & L, [3] 
	&M70 & [3,68] (M126 $\rightarrow$ L ) \\
{\bf M15} & C, [7,11] (M31 $\rightarrow$  [27])
	&{\bf M78} & L, XOR,$S_L$ (M118 $\rightarrow$ [27])\\
{\bf M16} & I, AND, $S_{L,A,AA}$
	&M84 & L, $S_{L,AA}$, [68]   \\
M17 & $S_L$,[1,16]
	&M85 & $S_L$ (M171 $\rightarrow$ [170])  \\
M18 & L, C, $S_A$, AND, [16,56]     
	&M86 & L  \\
M19 & L, C,[3,27] 
	&M94 &L,  $S_L$ (M102 $\rightarrow$ [11]) \\
M21 & $S_A$     
	&M97 &  $S_A$ \\
M23 & C, $S_A$,[7]             
	&M98 & $S_L$ \\
M26 &  $S_L$,[27]  
        &M99 &  \\
{\bf M27} &{\bf DD},C,AND,$S_L$,[11] (M79 $\rightarrow$ {\bf R+R},OR)
	&M101 &   \\
M28 & L          
	&M106 &  \\
M29 &                 
	&M108 & $S_{AA}$ (M124 $\rightarrow$ L) \\
M30 & L               
	&M113 & XOR, $S_A$ \\
M40 & AND, $S_{A,AA}$, [56] 
	&M114 & $S_L$ \\
M41 &  [3]                
	&{\bf M170} & I,XOR,$S_{L,A,AA}$,[186]\\
M42 &  $S_A$, [170]  
	&{\bf M186} & I,OR,$S_{L,A,AA}$,[170] (M325 $\rightarrow$ AND)\\
\hline
\end{tabular}
\end{center}
\caption{
{\bf 1L}: single-locus models (D: dominant, R: recessive, I and
$\overline{I}$: interference); 
{\bf RR}: jointly-recessive-recessive model;
{\bf DD}: jointly-dominant-dominant model;
{\bf RD}: jointly-recessive-dominant model;
{\bf T}: threshold model;
I: interference models.
L: missing lethal genotype models;
C: conditionally dominant and/or conditionally recessive; 
AND: logical AND models (multiplicative); 
OR: logical OR models (heterogeneity models); 
XOR: logical XOR models; 
S: symmetric models 
($S_L$: with respect to permutation of two loci; 
$S_A$: with respect to permutation of two alleles at one locus;
$S_{AA}$: with respect to permutation of two alleles at both loci);
[ ]: modifying-effect models. For example, [11] indicates a model that 
modifies M11 by one bit in the penetrance table.
}
\end{table}

\section{Marginal penetrance tables}

One important question we ask is how a two-locus model differs from
a single-locus model. This question has practical implications in 
linkage analyses because almost all current analyses are carried 
out by focusing on one susceptibility gene. We can
use the marginal penetrance table on each one of the two loci
to represent the effective single-locus model as the effects of
other interacting genes are averaged out. The marginal penetrance
table on the first locus is: $ f_i^{eff1} = \sum_{j} P_j^2 f_{ij} $ where 
$\{ P_j^2 \}$ are the genotype frequencies at the second locus, and 
that on the second locus is $f_j^{eff2}= \sum_{i} P_i^1 f_{ij}$, 
where $\{ P_i^1 \}$ are the genotype frequencies at the first locus. 

Take the modifying-effect model M15, for example. If $p_1$ and $p_2$ 
are disease allele frequencies at the two loci ($q_1=1-p_1, 
q_2=1-p_2$, and Hardy-Weinberg equilibrium is assumed), the
corresponding genotype frequencies are:
\begin{equation}
\begin{array}{cccc}
 &bb (q_2^2) &bB (2p_2q_2) & BB(p_2^2)\\
aa(q_1^2) & 0 & 0 & 0 \\
aA(2p_1q_1) & 0 & 0 & 1 \\
AA(p_1^2) & 1 & 1 & 1 \\
\end{array}
\end{equation}
The three marginal penetrances at the first locus are $(0, p_2^2, 1)$. 
As expected, it is very similar to the recessive model except for 
a modifying effect on the heterozygote. Similarly, the three 
marginal penetrances at the second locus are $(p_1^2, p_1^2, p_1^2 +2p_1q_1)$,
which are almost zero when $p_1$ is small. If linkage analysis for 
markers near both disease genes is carried out, the marker near the first gene
will provide a linkage signal under the recessive model with
a modified (reduced) penetrance; the marker near the second
gene will barely provide any linkage signal.

\begin{table}
\begin{tabular}{|c|cccc|cccc|c|cccc|cccc|}
\hline
model &\multicolumn{4}{c|}{first locus} & \multicolumn{4}{c|}{second locus} &
model &\multicolumn{4}{c|}{first locus} & \multicolumn{4}{c|}{second locus}\\
\cline{2-9} \cline{11-18}
\#& aa & aA & AA & type & bb & bB & BB & type
&\#&aa &aA &AA & type &bb &bB &BB & type \\ \hline
M1& 0 & 0  & .01 & - &0  & 0  & .01 & - &
	M43 & 0 & .82 & .19 & I & .18 & .01 & .19 & $\overline{I}$\\
M2 & 0 & 0  & .18 & R & 0& .01 & 0 & - &
	M45 & 0 & .82 & .82 &D & .19 & 0 & .19 & $\overline{I}$ \\
M3 & 0 & 0 & .19 & R & 0 & .01 & .01 & - &
	M56 & 0 & 1 & 0 & I & .18 & .18 & .18 & - \\
M5 & 0 & 0 & .82 & R & .01 & 0 & .01 & -  &
	M57 & 0 & 1 & .01 & I & .18 & .18 & .19 & - \\
M7 & 0 & 0 & 1 & R & .01 & .01 & .01 &  - &
	M58 & 0 & 1 & .18 & I & .18 & .19 & .18 & -  \\
M10 & 0 & .01 & .18 & R & 0 & .01 & .18 & R &
	M59 & 0 & 1 & .19 & I & .18 & .19 & .18 & -  \\
M11& 0 & .01 & .19 & R & 0 & .01 & .19 & R&
	M61 & 0 & 1 & .82 & D & .19 & .18 & .19 & -  \\
M12 & 0 & .01 & .81 & R & .01 & 0 & .18 & R &
	M68 & .01 & 0 & .81& R & .01 & 0 & .81 & R\\
M13 & 0 & .01 & .82 & R & .01 & 0 & .19 & R &
	M69 & .01 & 0 & .82 & R & .01 & 0 & .82 & R\\
M14 & 0 & .01 & .99 & R &.01 & .01 & .18 & R  &
	M70 & .01 & 0 & .99 & R & .01 & .01 & .81 & R \\
M15 & 0 & .01 & 1 & R &.01 & .01 & .19  & R  &
	M78 & .01 & .01 & .99 & R & .01 & .01 & .99 & R \\
M16 & 0 & .18 & 0 & I & 0 & .18 & 0 & I  &
	M84 & .01 & .18 & .81 & R & .01 & .18 & .81 & R \\
M17 & 0 & .18 & .01 & I & 0 & .18 & .01 & I &
	M85 & .01 & .18 & .82 & R & .01 & .18 & .82 & R \\
M18 & 0 & .18 & .18 & D & 0 & .19 & 0 & I  &
	M86 & .01 & .18 & .99 & R & .01 & .19 & .81 & R \\
M19 & 0 & .18 & .19 & D & 0 & .19 & .01 & I &
	M94 & .01 & .19 & .99 & R & .01 & .19 & .99 & R \\
M21 & 0 & .18 & .82 & R & .01 & .18 & .01 & I  &
	M97 & .01 & .81 & .01 & I  & .18 & 0 & .82 &  R \\
M23 & 0 & .18 & 1 & R  &.01 & .19 & .01 & I &
	M98 & .01 & .81 & .18 & I  & .18 & .01 & .81 & R  \\
M26 & 0 & .19 & .18 & D & 0 & .19 & .18 & D &
	M99 & .01 & .81 & .19 & I  & .18 & .01 & .82 & R\\
M27 & 0 & .19 & .19 & D  & 0 & .19 & .19 & D &
	M101 & .01 & .81 & .82 & D & .19 & 0 & .82 &R \\
M28 & 0 & .19 & .81 & R  &.01 & .18 & .18 & D &
	M106 & .01 & .82 & .18 & I  & .18 & .01 & .99 & R \\
M29 & 0 & .19 & .82 & R & .01 & .18 & .19 & D &
	M108 & .01 & .82 & .81 & D & .19 & 0 & .99 & R\\
M30 & 0 & .19 & .99 & R  &.01 & .19 & .18 & D  &
	M113 & .01 & .99 & .01 & I  & .18 & .18 & .82 & R \\
M40 & 0 & .82 & 0 & I & .18 & 0 & .18 & $\overline{I}$ &
	M114 & .01 & .99 & .18 & I  & .18 & .19 & .81 & R \\
M41 & 0 & .82 & .01 & I &.18 & 0 & .19 &  $\overline{I}$  &
	M170 & .18 & .82 & .18 & I  & .18 & .82 & .18 & I \\
M42 & 0 & .82 & .18 & I & .18 & .01 & .18 & $\overline{I}$  &
	M186 & .18 & 1 & .18 & I  & .18 & 1 & .18 & I \\
\hline
\end{tabular}
\caption{
Marginal penetrance tables at both loci for all $N_2-1=$50 two-locus models
assuming disease allele frequencies $p_1=p_2=0.1$. D,R,I,$\overline{I}$
represents (approximately) dominant, recessive, interference, and negative 
interference.  The symbol ``-" represents the case where the 
penetrance is not very sensitive to changes in the genotype.
}
\end{table}

\begin{table}
\begin{tabular}{|c|cccc|cccc|c|cccc|cccc|}
\hline
model &\multicolumn{4}{c|}{first locus} & \multicolumn{4}{c|}{second locus} &
model &\multicolumn{4}{c|}{first locus} & \multicolumn{4}{c|}{second locus}\\
\cline{2-9} \cline{11-18}
\#& aa & aA & AA & type & bb & bB & BB & type &
\# &aa &aA &AA & type &bb &bB &BB & type \\ \hline
M31  & 0 & .19 & 1 &  R &  .01 & .19 & .19  & D
 & M171 & .18 & .82 & .19 & I  &   .18 & .82 & .19 & I \\
M47 & 0 & .82 & 1 &  D  & .19 & .01 & .19 & $\overline{I}$
 & M173 & .18 & .82 & .82 &  D & .19 & .81 & .19 & I \\
M63 & 0 & 1 & 1 &  D &.19 & .19 & .19 & -
 & M175 & .18 & .82 & 1 &  D & .19 & .82 & .19 & I \\
M71 & .01 & 0 & 1 &  R  & .01 & .01 & .82 & R
 & M187 & .18 & 1 & .19 &  I & .18 & 1 & .19 &  I \\
M79  & .01 & .01 & 1 &  R &.01 & .01 & 1 & R
 & M189 & .18 & 1 & .82 &  D & .19 & .99 & .19 &  I \\
M87 & .01 & .18 & 1 &  R & .01 & .19 & .82 & R
 & M191 & .18 & 1 & 1 &  D &.19 & 1 & .19 &  I \\
M95  & .01 & .19 & 1 & R  &.01 & .19 & 1  & R
 & M229 & .19 & .81 & .82 & D  &  .19 & .81 & .82 &  D\\
M102 & .01 & .81 & .99 & D & .19 & .01 & .81 & R
 & M231 & .19 & .81 & 1 &  D & .19 & .82 & .82 &  D \\
M103 & .01 & .81 & 1 &  D & .19 & .01 & .82 & R
 & M238 & .19 & .82 & .99 &  D  &.19 & .82 & .99 &  D\\
M105 & .01 & .82 & .01 &  I & .18 & 0 & 1 &R
 & M239 & .19 & .82 & 1 &  D & .19 & .82 & 1 &  D\\
M107 & .01 & .82 & .19 &  I & .18 & .01 & 1 & R
 & M245 & .19 & .99 & .82 &  D  & .19 & .99 & .82 & D \\
M109 & .01 & .82 & .82 &  D  & .19 & 0 & 1 &R
 & M247 & .19 & .99 & 1 &   D & .19 & 1 & .82 &  D\\
M110 & .01 & .82 & .99 & D   & .19 & .01 & .99 & R
 & M254 & .19 & 1 & .99 &   D & .19 & 1 & .99  & D \\
M111 & .01 & .82 & 1 &  D & .19 & .01 & 1 & R
 & M255 & .19 & 1 & 1 &   D & .19 & 1 & 1  & D \\
M115 & .01 & .99 & .19 & I  & .18 & .19 & .82 & R
 & M325 & .82 & 0 & .82 &  $\overline{I}$ & .82 & 0 & .82 &  $\overline{I}$ \\
M117 & .01 & .99 & .82 & D  & .19 & .18 & .82 & R
 & M327 & .82 & 0 & 1 &  $\overline{I}$ & .82 & .01 & .82 &  $\overline{I}$\\
M118 & .01 & .99 & .99 & D & .19 & .19 & .81 & R
 & M335 & .82 & .01 & 1 & $\overline{I}$  & .82 & .01 & 1 &  $\overline{I}$ \\
M119 & .01 & .99 & 1 & D  &.19 & .19 & .82 & R
 & M341 & .82 & .18 & .82 &  $\overline{I}$ & .82 & .18 & .82 & $\overline{I}$\\
M121 & .01 & 1 & .01 & I & .18 & .18 & 1 & R
 & M343 & .82 & .18 & 1 & $\overline{I}$  & .82 & .19 & .82 & $\overline{I}$ \\
M122 & .01 & 1 & .18 & I & .18 & .19 & .99 & R
 & M351 & .82 & .19 & 1 &  $\overline{I}$ & .82 & .19 & 1 & $\overline{I}$ \\
M123 & .01 & 1 & .19 &  I & .18 & .19 & 1 & R
 & M365 & .82 & .82 & .82 &  - & 1 & 0 & 1 & $\overline{I}$ \\
M124 & .01 & 1 & .81 &  D &.19 & .18 & .99 & R
 & M367 & .82 & .82 & 1 &  - & 1 & .01 & 1 &  $\overline{I}$ \\
M125 & .01 & 1 & .82 &  D & .19 & .18 & 1 & R
 & M381 & .82 & 1 & .82 & - &  1 & .18 & 1 & $\overline{I}$ \\
M126 & .01 & 1 & .99 &  D & .19 & .19 & .99 & R 
 & M383 & .82 & 1 & 1 &  - & 1 & .19 & 1  &  $\overline{I}$ \\
M127  & .01 & 1 & 1 &  D & .19 & .19 & 1 & R
 & M495 & 1 & .82 & 1 & -  &   1 & .82 & 1 &  -  \\
\hline
\end{tabular}
\caption{
Similar to Table 4, but for $N_1-N_2-1=$50 two-locus models that
are equivalent to the models in Table 4 by switching the affection 
status and possibly other permutations between loci and alleles.
}
\end{table}

Assuming $p_1=p_2=0.1$, Table 4 lists the marginal penetrance
at both loci for all $N_2-1=$50 two-locus models. Table 5 lists those for 
the remaining $N_1-N_2-1=$50 models. Each marginal penetrance
on a single locus is roughly classified as one of the four types: 
dominant (D), recessive (R), interference (I), and negative 
interference ($\overline{I}$). Note that this classification
only provides crude guidance for marginal single-locus effect.
For example, in Table 4 the marginal penetrance table (0,0.2,0.8) is 
classified as recessive, though it is only approximately recessive with
some phenocopy probability.  Also note that for models that are 
equivalent to the representative models listed in Tables 3 and 4,
the marginal penetrances need to be recalculated using the correct 
allele frequencies.

Marginal penetrance tables can provide insight into linkage
analyses using a single-locus model when the underlying disease
model involves two genes. For example, for M1 (RR), both genes
behave like a recessive locus but with a highly reduced penetrance 
(0.01 if the disease allele frequency is 0.1). A single-locus-based
linkage analysis might detect both loci but with difficulty because
of the low penetrance.   M78 (an XOR model) provides another example.
It  is almost identical to M79 (R+R) in that both genes behave as a 
recessive locus, but the marginal penetrance is reduced from 1 to
0.99.  The almost negligible effect with the exclusive OR operation at 
the AA-BB genotype is due to the fact that the population frequency
of the AA-BB genotype is very small. In practice, it might be very 
difficult to distinguish M78 from M79 in a single-locus-based
linkage analysis. 

It is important to know that Tables 4 and 5 are derived with
a particular disease allele frequency ($p_1=p_2=$0.1). When
the disease allele frequency is the same as the normal allele
frequency ($p_1=p_2=$0.5), the nature of the marginal single-locus
model could be completely different. For example, the marginal
effect of both loci in M84 is between recessive and dominant when 
$p_1=p_2=$0.1. When $p_1=p_2=$0.5, the marginal penetrance becomes
(0.25, 0.5, 0.25) at both loci, similar to an interference model.
If the penetrance $f_{22}$ is 0.5 instead of 1, the marginal
penetrance is (0.25,0.25,0.25) \cite{frankel}; in other words,
there is no marginal linkage signal at all. 

In a practical pedigree analysis, the genotype frequencies
may not be taken from the population frequencies, but taken
from the pedigrees  one has \cite{vieland92,vieland93,sham94}.
It is thus possible that the penetrance table is specific
to each individual in the pedigree. It is another way of saying
that the risk of developing the disease for each family member 
is conditional on the affection status of other family members,
and such conditional probability may differ from person to
person.

\section{IBD probabilities in two-locus models}

There is a growing interest in using identity-by-descent (IBD)
sharing between affected sibpairs or affected relative pairs
to test whether a marker is linked to a susceptibility gene.
The premise behind the IBD test is that affected sib pairs
or affected relative pairs should share more IBD near the region
of the disease gene than expected from a random segregation. 
IBD sharing at one location is usually determined regardless of 
IBD sharing at other chromosomal locations, in order words, a
single-locus model is implicitly assumed. To test for possible
interactions between two regions, joint IBD sharing is needed
\cite{dizier86,risch90_2,knapp94,cordell95,farrall97}.

The observed joint IBD sharing can be compared with expected
IBD sharing under a certain model. There are at least three approaches 
in determining  the expected joint IBD sharing probability at two loci
between two affected sibs or affected relatives given a disease model. 
The first is to list all mating types, and count the number 
of each sharing situation among all possibilities. The second 
is to calculate the covariance of a quantitative trait between 
two relatives \cite{cockerham,kemp1,kemp2}. This covariance 
is decomposed into the sum of the products of ``coefficient 
of parentage" (or kinship coefficient) \cite{malecot} and the 
variance components. The latter includes additive and dominant
variance components by a linear regression of the quantitative
trait to the number of alleles \cite{fisher}. The conversion from
the covariance of a quantitative trait to the IBD sharing between
affected relatives can be accomplished by Bayes' theorem. The
third, and perhaps the more elegant approach, is to use Bayes' 
theorem to convert the probability of IBD sharing, given that the
two relatives are affected, to the probability of two relatives
being affected, given the IBD sharing. This approach was first
developed by Li and Sacks in 1954 \cite{lisacks,ccli_book}.

In Li-Sacks' original approach, a set of conditional probabilities,
the probability that the second relative has a certain genotype given
the first relative having a certain genotype, is conveniently written
in three 3-by-3 matrices (``Li-Sacks matrices") or four 4-by-4
matrices \cite{campbell}. These approaches were modified 
in \cite{wli98} by using two 2-by-2 matrices, which are the 
conditional probabilities that the second relative has a certain 
allele derived from one parent, given that the first relative  
has a certain allele derived from the same parent. In this
formulation, the probability that the two affected sibs share
$k1_m$ maternal alleles IBD and $k1_p$ paternal alleles IBD at
the first locus, and $k2_m$ maternal alleles IBD and $k2_p$ 
paternal alleles IBD at the second locus is
$$
P(k1_m,k1_p,k2_m,k2_p|\mbox{both sibs affected}) = 
\frac{\mbox{numerator N}}{\mbox{denominator D}}
$$
with
\begin{eqnarray}
N& = & \sum_{i1_m,i1_p,i2_m,i2_p,j1_m,j1_p,j2_m,j2_p} f_{j1_mj1_pj2_mj2_p} 
\cdot
t_{i1_mj1_m}(k1_m) t_{i1_pj1_p} (k1_p) t_{i2_mj2_m}(k2_m) t_{i2_pj2_p}(k2_p)
\nonumber \\
& & \cdot f_{i1_mi1_pi2_mi2_p}
 p_{i1_m} p_{i1_p} p_{i2_m} p_{i2_p}
\cdot p(k1_m) p(k1_p) p(k2_m) p(k2_p) 
\nonumber \\
D &= &  \mbox{(sum of N over $k1_m,k1_p,k2_m,k2_p$)} 
\end{eqnarray}
where
\begin{itemize}
\item
$i1_m$ is the index for the maternally derived allele
(the paternally derived allele uses the label p), in the first sib
(second sib uses the label $j$), at the first locus (second locus 
uses the label 2) 

\item
$f_{i1_mi1_pi2_mi2_p}$ and $f_{j1_mj1_pj2_mj2_p}$ are the
penetrance tables of the two-locus model. Although it has 4 indices,
it can be easily obtained from the 3-by-3 penetrance table as in
Eq.\ref{eq1}. 

\item
$p_{i1_m}, p_{i1_p}, p_{i2_m}, p_{i2_p}$ are the allele 
frequencies, which take the value of either $p_1$ or $q_1=1-p_1$.

\item
$p(k1_m), p(k1_p), p(k2_m), p(k2_p)$ are the prior probabilities
of sharing allele IBD at four places (maternally and paternally
derived, first and second locus), which are 1/2's for 
sibpairs.

\item
$t_{i1_mj1_m}(k1_m), t_{i1_pj1_p}(k1_p), t_{i2_mj2_m}(k2_m), 
t_{i2_pj2_p}(k2_p)$ are the revised 2-by-2 Li-Sacks matrices given by:
\begin{equation}
\{ t_{ij}(1) \} = \left( \begin{array}{cc} 1 & 0 \\ 0 & 1\end{array} \right),
\{ t_{ij}(0) \} = \left( \begin{array}{cc} p & q \\ p & q\end{array} \right)
\end{equation}
\end{itemize}
Despite the complicated indexing, the revised Li-Sacks approach is
easier to implement in a computer code, and easier to generalize 
to other situations, such as unilineal relative pairs, multiple 
alleles, unaffected-unaffected and unaffected-affected pairs, 
the probability of identity-by-state, two markers instead of 
two disease genes, etc. \cite{wli98}. More details will be 
discussed elsewhere [Li, in preparation].

There are two types of joint IBD measurements currently in use: the 
first is the addition of maternal and paternal IBDs, which take the values 
of 0,1,2:
\begin{equation}
P_{geno}(k1,k2)=
\sum_{k1=k1_m+k1_p, k2=k2_m+k2_p} P(k1_m,k1_p,k2_m,k2_p). 
\end{equation}
The genotypic IBD's, $\{ P_{geno}(k1,k2) \}$, form a 3-by-3 matrix. The second 
measurement focuses on maternal (or equivalently, paternal) 
IBD only:
\begin{equation}
P_{alle}(k1_m,k2_m)=  
\sum_{k1_p,k2_p} P(k1_m,k1_p,k2_m,k2_p).
\end{equation}
The symmetry between the maternally-derived and paternally-derived
alleles implies that $P(k1_p,k2_p)= P(k1_m,k2_m)$.
The allelic IBD's, $\{ P_{alle}(k1_m,k2_m) \}$, form a 2-by-2 matrix, 
which will be the joint IBD measurement we use. For example, 
for M15 at $p_1=p_2=0.1$, the joint allelic IBD is:
\begin{equation}
\label{eq9}
\begin{array}{cccc}
 & k2_m=0 &k2_m=1 & \mbox{marginal $k1_m$}  \\
k1_m=0 & 0.050549 & 0.072689 & 0.123238  \\
k1_m=1 & 0.413962 & 0.462800 & 0.876762  \\
\mbox{marginal $k2_m$} &0.464511 & 0.535489 & 1
\end{array}
\end{equation}
The marginal probabilities of IBD sharing in Eq.\ref{eq9} confirms 
our intuition that there is a strong preference for the IBD sharing 
on the first locus to be 1 (probability of sharing 0.876762 versus 
non-sharing 0.123238), whereas the deviation from 0.5 at the second 
locus is very small (0.535489 versus 0.464511).

\section{Correlation between IBD sharings at two loci}

For probabilities of joint IBD sharings at two loci as exemplified
by Eq.\ref{eq9}, we ask the following question: Can the joint probability 
be derived from the two marginal IBD sharing probabilities at the 
two separated loci? This question is motivated by the suggestion in 
\cite{maclean,cox} that one might first detect marginal effects by
single-locus linkage analysis, then detect interaction later using
the correlation analysis. Such a correlation between two marginals exists
only if the joint probability is not equal to the product of the
two marginals.  Statistical correlations can be measured in 
different ways, one of them being the mutual information, defined 
as\cite{kullback,wli90_1}:
\begin{equation}
\label{mutu}
M = \sum_{k1_m, k2_m} P(k1_m,k2_m) 
\log_2 \frac{P(k1_m,k2_m)}{P(k1_m, \cdot)P(\cdot, k2_m)}
\end{equation}
where $P(k1_m, \cdot)$ and $P(\cdot, k2_m)$ are the
two marginal IBD sharing probabilities at two loci. Mutual information 
has certain meaning in information theory, and is intrinsically 
related to the concept of entropy. Two is chosen as the base of 
the logarithm so that it is measured by the unit of ``bit", though
base $e$ and base 10 can also be used.

We calculate the mutual information for the 2-by-2 joint probabilities 
of allelic IBD sharing at two loci for all 50 two-locus models, at 3
different allele frequency values: $p_1=p_2=$0.001, 0.01, and 0.1. 
Also shown is an asymmetric situation when $p_1=0.1$ and $p_2=0.01$.
The result is summarized in Table 6 (and Table 7 for the other 50 models).
Only one significance digit is kept in Tables 6 and 7.

\begin{table}
\begin{tabular}{|c|cccc|c|cccc|}
\hline
model &\multicolumn{4}{c|}{disease allele freq} &
model &\multicolumn{4}{c|}{disease allele freq}\\
\cline{2-5} \cline{7-10}
number & 0.001& 0.01& 0.1& 0.1,0.01 & 
number &  0.001& 0.01& 0.1 &0.1,0.01 \\ \hline
M1$^*$ & 0 & 0 &  0 & 0  &
 M43 & 9e-14(P) & 9e-10(P) &  8e-6(P) & 8e-8(P) \\
M2$^*$ & 0 & 0 & 0 & 0 &
 M45$^*$ & 0 & 0 & 0 & 0\\
M3$^*$ & 0 & 0 & 0 & 0 & 
 M56$^*$ & 0 & 0 & 0 & 0\\
M5$^*$ & 0 & 0 & 0 & 0 &
 M57 & 0.0(P) & 0.0(P) & e-9(P) & e-13(P) \\
M7$^*$ & 0 & 0 & 0 & 0 &
 M58 & 0.0(P) & 4e-11(P) & 2e-7(P) & 4e-9(P) \\
M10 & 0.02(N) &  0.01(N) &  2e-4(N) &2e-4(N) & 
 M59 & 0.0(P) & 4e-11(P) & 2e-7(P) & 4e-9(P) \\
M11 & 0.02(N) &  0.01(N) & 9e-4(N) & 3e-4(N) &
 M61 & 0.0(P) & 3e-11(P) & 2e-7(P) & 3e-9(P) \\
M12 & 4e-5 (N) & 3e-4(N) & e-3(N) &2e-7(N) &
 M68 & 0.1(N) & 0.1(N) &  0.02(N) & 5e-5(N) \\
M13 & 4e-5(N) & 3e-4(N) & 2e-3(N) &3e-7(N) &
 M69 & 0.1(N) & 0.1(N) & 0.02(N) & 5e-5(N) \\
M14 & 4e-5(N) & 3e-4(N) & 8e-4(N) &2e-7(N) &
 M70 & 0.1(N) & 0.1(N) & 0.02(N) & 4e-5(N) \\
M15 & 4e-5(N) & 3e-4(N) & e-3(N) &3e-7(N) &
 M78 & 0.1(N) & 0.1(N)  & 0.03(N)  & 9e-5(N) \\
M16$^*$  & 0 & 0 & 0 & 0 &
 M84 & 9e-3(N) & 6e-3(N) & 9e-6(N) & e-3(N) \\
M17 & 2e-14(P) & 2e-10(P) & 2e-6(P) &2e-8(P) &
 M85 & 9e-3(N) & 6e-3(N) & e-5(N) & e-3(N) \\
M18$^*$ & 0 & 0 & 0 & 0 &
 M86 & 9e-3(N) &  7e-3(N) & 2e-4(N) & 2e-3(N) \\
M19 & 0.0(P) & 7e-11(P) &  2e-7(P) &3e-9(P)  &
 M94 & 9e-3(N) & 7e-3(N) & 4e-4(N)  & 2e-3(N) \\
M21 & 2e-3(N) & e-3(N) & 2e-5(P) & e-3(N) &
 M97 & e-8(N) & e-6(N) & e-5(N)  & 7e-10(N) \\
M23 & 2e-3(N) & 2e-3(N) & 7e-5(N) &2e-3(N)  &
 M98 & e-8(N) & e-6(N) & 5e-8(N)  & 6e-8(P) \\
M26 & 0.0(N) & 2e-12(N) & 5e-9(P) & 3e-13(P) & 
 M99 & e-8(N) & e-6(N) & 6e-8(N)  & 6e-8(P) \\
M27$^*$  & 0 & 0 & 0 & 0 &
 M101 & e-8(N) & e-6(N) & 5e-6(N) & 3e-10(N) \\
M28 & 2e-3(N) & e-3(N) &  3e-5(P) & e-3(N) &
 M106 & e-8(N) & e-6(N) &  e-5(N) & 5e-8(P) \\
M29 & 2e-3(N) & e-3(N) & 2e-5(P) & e-3(N) &
 M108 & e-8(N) & e-6(N) & 3e-5(N) & 2e-9(N) \\
M30 & 2e-3(N) & 2e-3(N) & 4e-5(N) & 2e-3(N) &
 M113 & e-8(N) & e-6(N) & 5e-6(N) & 6e-10(N)\\
M40$^*$ & 0 & 0 & 0 & 0 &
 M114 & e-8(N) & e-6(N) & 2e-6(N) & e-9(P)\\
M41 & 0.0(P) & 0.0(P) &  8e-10(P) & e-13(P) &  
 M170 & 3e-3(N) & 2e-3(N) &  7e-5(P) & e-5(N) \\
M42 & 9e-14(P) & 9e-10(P) & 8e-6(P) & 8e-8(P) &
 M186 & 3e-3(N) & 2e-3(N) & e-4(N) & 7e-5(N) \\
\hline
\end{tabular}
\caption{
Values of mutual information (with one significance digit) between 
the two marginal probabilities of IBD sharing for all 
$N_2-1=$50 two-locus
models. The allele frequencies are chosen at four different
values:  $p_1=p_2=$0.001, 0.01, 0.1; $p_1=0.1$ and $p_2=0.01$.
Values lower than $10^{-14}$ are converted to 0. ``4e-5" means
to $4 \times 10^{-5}$, etc. Multiplicative models are marked by $^*$.
}
\end{table}

\begin{table}
\begin{tabular}{|c|cccc|c|cccc|}
\hline
model &\multicolumn{4}{c|}{disease allele freq} &
model &\multicolumn{4}{c|}{disease allele freq}\\
\cline{2-5} \cline{7-10}
number & 0.001& 0.01& 0.1& 0.1,0.01 & 
number& 0.001& 0.01& 0.1 & 0.1, 0.01 \\ \hline
M31 & 2e-3(N) & 2e-3(N) &  4e-5(N) & 2e-3(N) &
 M171 & 3e-3(N) & 2e-3(N) & 7e-5(P) &e-5(N)  \\
M47 & 3e-14(P) & 3e-10(P) & e-6(P) & e-8(P) & 
 M173 & 3e-3(N) & 2e-3(N) & 8e-5(P) & 6e-6(N) \\
M63$^*$ & 0 & 0 & 0 & 0 &
 M175 & 3e-3(N) & 2e-3(N) & 7e-5(P) & 6e-6(N) \\
M71 & 0.1(N) & 0.1(N) & 0.03(N)  & 4e-5(N) & 
 M187 & 3e-3(N) & 2e-3(N) & e-4(N)  & 7e-5(N) \\
M79 & 0.1(N) & 0.1(N) & 0.03(N) & 9e-5(N) &
 M189 & 3e-3(N) & 2e-3(N) & 7e-5(N) & 4e-5(N) \\
M87 & 9e-3(N) &  7e-3(N)  &  2e-4(N) & 2e-3(N) &
 M191 & 3e-3(N) & 2e-3(N) & 8e-5(N) & 4e-5(N) \\
M95 & 9e-3(N) & 7e-3(N) & 4e-4(N) & 2e-3(N) &
 M229 & 3e-3(N) & 2e-3(N) & 9e-5(P)  & 6e-6(N)\\
M102 & e-8(N) & e-6(N)  & e-6(N) & e-8(P) &
 M231 & 3e-3(N) & 2e-3(N)  &  8e-5(P)  & 6e-6(N) \\
M103 & e-8(N) & e-6(N)  & e-6(N) & e-8(P) &
 M238 & 3e-3(N) & 2e-3(N)  & 7e-5(P) & 6e-6(N) \\
M105 & e-8(N) & e-6(N)  & 5e-5(N) & 3e-9(N) &
 M239 & 3e-3(N) & 2e-3(N)  & 7e-5(P) & 6e-6(N) \\
M107 & e-8(N) & e-6(N)  & e-5(N) & 5e-8(P) &
 M245 & 3e-3(N) & 2e-3(N)   & 4e-5(N)  &  4e-5(N) \\
M109  & e-8(N) & e-6(N) & 3e-5(N) & 2e-9(N) &
 M247 & 3e-3(N) & 2e-3(N)  & 5e-5(N)  & 4e-5(N) \\
M110 & e-8(N) & e-6(N)  & 2e-5(N) & 6e-9(P) &
 M254 & 3e-3(N) & 2e-3(N)  & 6e-5(N)  & 4e-5(N) \\
M111 & e-8(N) & e-6(N) & 2e-5(N) & 5e-9(P)  &
 M255 & 3e-3(N) & 2e-3(N) & 7e-5(N) & 4e-5(N) \\
M115 & e-8(N) & e-6(N) &  2e-5(N) & e-9(P) &
 M325$^*$ & 0  & 0  & 0  & 0 \\
M117 & e-8(N) & e-6(N) & 1e-6(N)  & 2e-9(P) &
 M327 & 0(P) & e-14(P) & 8e-9(P) & e-10(P) \\
M118 & e-8(N) & e-6(N) & 2e-6(N) & 3e-10(N) &
 M335 & 0(P)  & 4e-14(P)  & 3e-8(P) & e-10(P) \\
M119 & e-8(N) & e-6(N) & 2e-6(N)  & 3e-10(N)  &
 M341 & 4e-13(P) & 4e-9(P) & 4e-5(P) & 4e-7(P) \\
M121  & e-8(N) & e-6(N) & 2e-5(N) & 3e-9(N) &
 M343 & 4e-13(P)  & 4e-9(P)  & 4e-5(P) & 4e-7(P) \\
M122 & e-8(N) & e-6(N) & 2e-5(N)  & 3e-11(P) &
 M351 & 4e-13(P)  & 4e-9(P)  & 4e-5(P) & 4e-7(P) \\
M123 & e-8(N) & e-6(N) & 2e-5(N)  & 2e-11(P) &
 M365$^*$ & 0  & 0  & 0  & 0 \\
M124 & e-8(N) & e-6(N) & e-5(N)  & 2e-10(P) &
 M367 & 0(P) & e-14(P) & 8e-9(P) & e-10(P) \\
M125 & e-8(N) & e-6(N) & e-5(N) & 2e-10(P) &
 M381 & 4e-14(P) & 4e-10(P) & 2e-6(P) & 2e-8(P)  \\
M126 & e-8(N) & e-6(N) & e-5(N)  & 2e-9(N) &
 M383 & 4e-14(P) & 4e-10(P)  &  2e-6(P) &3e-8(P)  \\
M127 & e-8(N) & e-6(N) & 7e-5(N)  & 2e-9(N)  &
 M495 &  4e-14(P)  & 4e-10(P)  & e-6(P) &2e-8(P) \\
\hline
\end{tabular}
\caption{
Similar to Table 6 but for the $N_1-N_2-1=$50 models that are 
equivalent to the models in Table 6 by switching affection status.
}
\end{table}

Table 6 confirms the conclusion in \cite{hodge81_1} that for 
multiplicative models, the IBD sharing probability at one locus can 
be calculated as if there is no interaction with another locus:
the correlation as measured by mutual information is 0 for all 
these models. 

It should be of interest to examine which two-locus models
exhibit the smallest correlation, and which the largest.
Besides the zero correlation for multiplicative and single-locus
models, all modifying-effect models as altered from a single-locus
model or a multiplicative model should exhibit small correlations.
Indeed, in Table 6, we see that at $p_1=p_2=0.001$, M19, M26, M41, M57, 
M58, M59, M61 all exhibit close-to-zero correlations.

From Tables 6 and 7, it seems that missing lethal genotype models 
tend to have larger correlation values, although these values are
derived from a limited choice of parameter settings.  To some extent, 
this observation is not surprising. Missing lethal genotype 
models are typically ``non-linear" in the sense that as the 
sum of the total number of disease alleles is increased, the 
change in phenotype is not monotonic (it can first change from 
unaffected to affected, then from affected to unaffected). For 
these models, using the joint IBD sharing probability to detect 
linkage should have the greatest increase of power over methods 
using marginal probability of IBD sharing.

Occasionally, not only would we like to know the ``strength" or
``magnitude" of the correlation between the marginal IBD sharing  
probabilities at two loci, but also the sign of the correlation. 
For example, in \cite{maclean,cox}, whether the statistical 
correlation between two linkage signals obtained at two loci is 
positive or negative provides an indication as to whether the two
loci are ``interacting" or simply heterogeneous. We provide this
piece of information for all two-locus models in Tables 6 and 7.
A ``(P)" indicates that $P(k1_m=1, k2_m=1)$ is larger than
the expected value from no correlation $P(k1_m=1) \cdot P(k2_m=1)$;
similarly, an ``(N)" indicates that the joint probability is smaller
than the product of two marginals. As expected, all heterogeneity
models (M79,M127,M255) have negative correlations.

Note that we measure the correlation by a probability-based quantity 
rather than a statistics-based one. This is because we start 
with a theoretical model, i.e. a two-locus model, and investigate 
the consequence of the model. On the other hand, if we start with 
a sample of size $N$ and the count of joint IBD status $ij$ 
is $N_{ij}$ ($\sum_{ij} N_{ij}=N$), we can use any one of statistics
to test the significance of the correlation; for example,
the likelihood-ratio statistic, 
\begin{equation}
G^2 = 2N \sum_{ij} \frac{N_{ij}}{N} \log \frac{N_{ij}N}{N_{i.} N_{.j}},
\end{equation}
and the Pearson chi-square statistic,
\begin{equation}
X^2 = \sum_{ij} \frac{ (N_{ij}-N_{i.} N_{.j}/N )^2 }{N_{i.} N_{.j}/N},
\end{equation}
where $N_{i.} \equiv \sum_j N_{ij}$ and $N_{.j} \equiv \sum_i N_{ij}$
are the two marginal counts.  It can be shown (see Appendix 2) 
that $G^2$ and $X^2$ are approximately equal.  Under the no-correlation 
null hypothesis, both $G^2$ and $X^2$ approximately follow the $\chi^2$ 
distribution with 1 degree of freedom. The larger the $G^2$ 
and $X^2$, the more likely that the null hypothesis is wrong.

It is important to note that if the null hypothesis is indeed
incorrect,  both $G^2$ and $X^2$ increase with the sample size N.
Consequently, $G^2$ and $X^2$ do not measure the strength of
the correlation, but the evidence that no-correlation hypothesis is
wrong. On the other hand, the normalized quantities
such as $\sqrt{G^2/N}$ and $\sqrt{X^2/N}$ (``phi coefficient", 
page 741 of \cite{sokal}. or Cramer's V, page 631 of \cite{press})
do measure the correlation strength.  Compared with the mutual 
information defined in Eq.\ref{mutu}, we see that
$G^2/N \approx 2\log(2) M$.

\section{Discussions}

We present a complete enumeration and an attempt at classification
of 512 two-locus two-allele fully-penetrant disease models.
Excluding zero-locus and single-locus models, the minimum set
of non-redundant two-locus models is 48, and with the two single-locus
models included, 50. Even though the permutation of affection
status does not change the ``nature" of the interaction between
two genes, for many practical applications, it is helpful
to keep 50 other models which are equivalent to the first
50 models by this permutation in the penetrance table (plus
possibly other permutations between alleles and loci). For example,
a logical OR model (heterogeneity model) is equivalent to
a logical AND model (multiplicative model). Nevertheless,
the special property for a multiplicative model, that the
joint IBD sharing probability is equal to the product of two 
marginal IBD probabilities, does not hold for a heterogeneity
model. Even with our total 100 non-redundant models, the permutations
between alleles or loci require a corresponding change of allele 
frequencies in some calculations.

One of the main purposes of this paper is to point out that besides 
6 two-locus disease models typically used in linkage analysis 
assuming two interacting genes, there are many other types of 
gene-gene interactions. On one hand, we admit that many of the 
two-locus models may not describe a real interaction between two gene 
products in a genetic disease; on the other hand, it is fairly 
straightforward to construct a biochemical system based 
on a two-locus model. A prototypical biochemical system 
consists of proteins formed by one peptide, dimer proteins
formed by two complementary peptides, and dimer proteins
formed by two identical peptides. By specifying the functional and
non-functional proteins as well as the level of protein concentration
required by a normal phenotype, it is possible to materialize 
any two-locus models.

The marginal penetrance table we calculated in this paper is relevant 
to linkage analysis using only single-locus models. There have
been discussions of whether single-locus models are sufficient to
detect a linkage signal even if the underlying disease model may involve
gene-gene interaction 
\cite{green89,green90,goldin92,vieland92,goldin93,vieland93,sham94,schork94,olson95,dizier96,hodge97,todorov97}. 
Part of the answer can be predicted by the marginal penetrance table:
if the marginal penetrance table is clearly dominant or recessive, 
it is possible that a single-locus model is able to detect linkage;
otherwise, two-locus models should offer more power. Although it was 
mentioned that the gain of the logarithm of likelihood ratio (same as
log-of-odd, or LOD scores) by using two-locus models over those by 
single-locus models may be at most 17\% \cite{sham94}, after removing 
the logarithm,  the increase of the likelihood ratio can be much larger. 
For example, if the LOD score equals to 2, or the likelihood ratio is 
equal to 100, an increase in LOD of 17\% is equivalent to an increase 
in likelihood ratio of 118\%! What is considered as ``more" powerful
versus ``slightly more" powerful is not specified.

As a compromise between detecting linkage signals using single-locus
models and using two-locus models, it is suggested that a pairwise
correlation between linkage signals obtained by single-locus models
can be used to detect linkage for interacting genes \cite{maclean,cox}. 
A similar idea for detecting higher-order correlations among
linkage signals from different locations using artificial
neural networks is discussed in \cite{gaw11}. Our result 
on the sign and strength of correlation between two
marginal IBD sharing probabilities (Tables 6 and 7) is directly
relevant to this approach. We observed that models modified from 
the multiplicative and single-locus models exhibit a very weak
correlation, whereas missing lethal genotype models or ``non-linear"
models exhibit the strongest correlation. Since many two-locus
models share similar correlation values, of sign and magnitude,
we may not be able to distinguish them using this approach.

There are many topics on two-locus disease models that
are not discussed here. Some classification schemes discussed
in \cite{wli97} are not included (e.g. models that are
conditionally dominant or recessive with respect to two loci),
as well as the idea of genotype-induced representation of joint IBD 
distributions (Reich, unpublished results), and the idea of ``phase 
transition" in the two-locus model space (Li, unpublished results). 
The extension from fully-penetrant
models to reduced-penetrant models as well as models for quantitative 
traits is very important since many complex diseases are not
dichotomous.  Many calculations presented in this paper are 
implemented in a computer program: {\sl u2}  for ``utility program 
for two-locus models". More information on this program can be found 
at the web page {\sl http://linkage.rockefeller.edu/soft/u2}.

\hspace{0.5in}

\section*{Appendices}

\subsection*{1. A formal derivation of the value of $N_2$ by
de Bruijn's theorem}

Let's consider two permutations applied on the phenotypes: 
the identity operation and the exchange permutation. The cycle 
index of this permutation group on the phenotype is: 
$$
C_{pheno}(x_1, x_2) = \frac{x_1^2 + x_2}{2}.
$$

By de Bruijn's generalization of P\'{o}lya's theorem (theorem 5.4 in 
\cite{debruijn}), when the permutation group on phenotypes is 
considered, the number of equivalence two-locus models can be 
obtained by the following procedure: replacing $x_1$ in $C_{geno}$ by
the partial derivative $\partial /\partial x_1$, $x_2$
by  $\partial /\partial x_2$, etc., and applying the partial
derivative to $C_{pheno}$ while replacing $x_1$ with
$e^{(x_1+x_2+ \cdots)}$, $x_2$ with
$e^{2(x_2+x_4 + \cdots)}$, etc., then evaluating the
expression at $x_1=x_2 = \cdots = 0$ :
\begin{eqnarray}
N_2 &=&
\frac{1}{8} \left[
\frac{\partial^9}{\partial x_1^9} +
4 \frac{\partial^3}{\partial x_1^3} \frac{\partial^3}{\partial x_2^3} +
\frac{\partial}{\partial x_1} \frac{\partial^4}{\partial x_2^4} +
2 \frac{\partial}{\partial x_1} \frac{\partial}{\partial x_4}
\right] \nonumber \\
&&
\left.
\frac{1}{2}
\left[
e^{2(x_1 +x_2 +x_3 +x_4)} +e^{2(x_2+x_4)}
\right] \right|_{x_1=\cdots \cdots=0}   \nonumber \\
&=& 51. \nonumber
\end{eqnarray}

Since the permutation group on the phenotype considered here
is particularly simple, $N_2$ is simply $N_1$ divided by 2.

\subsection*{2. Approximate equivalence between $G^2$ and $X^2$}

If we write $J_{ij}=N_{ij}/N$, 
$S_{ij}= N_{i.}N_{.j}/N^2$, and assume the difference between the
two is small: $\Delta_{ij} \equiv J_{ij}-S_{ij}$, the following
approximation by a Taylor expansion, 
\begin{eqnarray}
2 \sum_{ij} J_{ij} \log \frac{J_{ij}}{S_{ij}}
& \approx &  2\sum_{ij} (S_{ij} + \Delta_{ij}) 
\log(1 + \frac{\Delta_{ij}}{S_{ij}})
\approx 2\sum_{ij} (S_{ij} + \Delta_{ij}) 
(\frac{\Delta_{ij}}{S_{ij}} - \frac{\Delta_{ij}^2}{2 S_{ij}}) \nonumber \\
&\approx& 
2 \sum_{ij} \Delta_{ij} + \sum_{ij} \frac{\Delta_{ij}^2}{S_{ij}} 
\approx
\sum_{ij} \frac{\Delta_{ij}^2}{S_{ij}} 
= \sum_{ij} \frac{ (J_{ij}-S_{ij})^2}{ S_{ij}},
\end{eqnarray}
shows that $G^2$ and $X^2$ are approximately equal \cite{agresti}. 

\section*{Acknowledgements}
This paper is expanded from the poster presented at the 1997 Annual 
Meeting of American Society of Human Genetics  \cite{wli97}. The
material in section 5 is an abridged version of the poster
presented at the 1998 Annual Meeting of American Society of 
Human Genetics  \cite{wli98}.  W.L. would like to thank Marcella 
Devoto for suggesting the topic on maternal-fetal incompatibility, 
Cathy Falk, Fatemeh Haghighi, Harald G\"{o}ring for commenting 
on a first draft, and Katherine Montague for proofreading the paper. 
W.L. is supported by the grant K01HG00024 from NIH and partial 
support from HG00008.


\end{document}